# On The Expansion and Fate Of The Universe

Aldo Bonasera[1,2]

*1) Laboratori Nazionali del Sud, INFN, via Santa Sofia, 62, 95123-Catania, Italy.*

*2) Cyclotron Institute, Texas A&M University, College Station, TX-77843, USA.*

*ABSTRACT*

The evolution of the universe from an initial dramatic event, the Big-Bang, is firmly established. Hubble's law [1] (HL) connects the velocity of galactic objects and their relative distance: v(r)=Hr, where H is the Hubble constant. In this work we suggest that HL is not valid at large distances because of total energy conservation. We propose that the velocity can be expanded in terms of their relative distance and produce a better fit to the available experimental data. Using a simple 'dust' universe model, we can easily calculate under which conditions an (unstable) equilibrium state can be reached and we can estimate the values of the matter present in the universe as well as the 'dark energy'. We do not need to invoke any 'dark energy', its role being played by the kinetic correction. The resulting picture is that the universe might reach an unstable equilibrium state whose fate will be decided by fluctuations: either collapse or expand forever.

Modern cosmology has reached such a very high precision to put some firm constraint on the birth and evolution of the universe [1-3]. The universe is evolving from an initial Big-Bang and it seems to be isotropic and homogeneous. This is supported by the Hubble's law (HL), which remains the same for two observers located in two different places in the universe [3]. In fact if $v_B$ and $v_A$ are the velocities of a given galaxy as measured by two observers located in two different galaxies A and B:

$$v_{AB} = v_B - v_A = H(r_B - r_A) = Hr_{AB} \qquad (1)$$

However, measurements [2,3] from the Wilkinson Microwave Anisotropy Probe (WMAP) collaboration, confirm the possibility of small temperature fluctuations, thus any deviation from HL must be small.

We briefly summarize the main experimental results relevant for this work. The matter density of the universe is given by [2,3]:

$$[\Omega_m]_{WMAP} = \frac{\rho_m}{\rho_c} = \frac{8\pi G \rho_m}{3H^2} = 0.27 \pm 0.04 \qquad (2)$$

Where $\rho_m$ is the matter density, which includes baryonic as well as *dark matter* (DM). The critical density $\rho_c = \frac{3H^2}{8\pi G}$, depends on the value of the Hubble constant H and G is Newton's gravitational constant. The baryonic density is estimated to be about 16% of the total matter density, the rest is matter in some other form. The value of the baryonic density is inferred from the observed primordial abundance of elements formed in the early universe [3-5]. The total matter density in equation 2 is obtained from gravitational effects on the motion of stellar objects. We will not discuss any further about this distinction and in our work we will derive the total density including dark and non-dark matter. A contribution to the density is due to relativistic particles, especially neutrinos and photons; their contribution is presently small and we will neglect it [2,3].

Observations and detailed calculations reveal that some form of energy is missing which suggested an 'ad hoc' potential term:

$$U_\Lambda = -\frac{1}{6}\Lambda mc^2 r^2 \qquad (3)$$

Where $\Lambda$ is the cosmological constant first introduced by A. Einstein [3]. Even though we are using Newtonian cosmology in equation 3 and in the following, the results remain unchanged when general relativity is used. The cosmological constant was first introduced with the desire of obtaining stationary solutions for the universe, a concept that turned out to be incorrect after Hubble's observations. Nevertheless, experimental data forced to include a form of unknown energy exemplified by equation 3 and dubbed as: *Dark Energy* (DE). The experimental value of the DE density, and implicitly of the cosmological constant $\Lambda$, is [2]:

$$[\Omega_\Lambda]_{WMAP} = \frac{\rho_\Lambda}{\rho_c} = \frac{\Lambda c^2}{3H^2} = 0.73 \pm 0.04 \qquad (4)$$

Equation 2 and 4 give the total density (neglecting the small contribution from relativistic particles):

$$[\Omega_0]_{WMAP} = [\Omega_\Lambda]_{WMAP} + [\Omega_m]_{WMAP} = 1.00 \pm 0.08 \qquad (5)$$

The model described above is known as the $\Lambda$CDM: it includes the cosmological constant $\Lambda$ and DM [3]. Since the $\Lambda$ constant is positive, equation 3 results in an acceleration: galaxies might be accelerating from each other at large distances, instead of decelerating, as one would naively expect from Newtonian's gravitational law. Most of nowadays theoretical and experimental efforts are concentrating on determining the nature of DM and DE. In this paper we will show that a simple modification of Hubble's law (MHL), can explain the experimental results without the need of invoking any DE as given in equation 3.

There is no reason why Hubble's law given in equation 1 should be valid for any value of the distance r. In particular it is already assumed in equation 1 that when r→0, v→0. A constant term could be added which might not be in contrast with actual experimental data. Such a term is usually neglected, in agreement with equation 1, and we will conform to such an assumption hoping for new data at smaller distances. The major problem arises when distances become very large. Physically, we do not expect velocities to diverge linearly. In particular, because of total energy conservation, we expect HL to be modified at large r. To take these features into account we propose an expansion of the velocity as:

$$v(r) = H_0 r - H_1 r^2 + O(r^3) \qquad (6)$$

In **fig. 1**, we plot the velocity of galaxies versus their relative distances obtained from the WMAP collaboration [2], full dots. In the same figure we display a linear fit according to HL (equation 1-full line), and a second order expansion (equation 6-dashed line). Clearly the MHL reproduces the data better in particular at large distances. The inclusion of even higher order terms is not justified by the available data, but it might be important if measurements at larger (as well as smaller) distances will be available in the future. Notice the small difference between the values of the constants H and $H_0$ in the two approximations. The value of $H_1$ is relatively small, and so is the correction to HL, which suggests that equation 1 is valid in first approximation, thus we expect fluctuations to be small in agreement to data. From the value of the constants in the MHL, we can define some typical times, distances and velocities:

$$\begin{cases} t_{MHL} = \frac{1}{H_0} = 4.2 \; 10^{17} s \\ l_{MHL} = \frac{H_0}{H_1} = 1.04 \; 10^{26} m \\ v_{MHL} = \frac{l_H}{t_H} = 2.5 \; 10^8 \; m/s \end{cases} \quad (7)$$

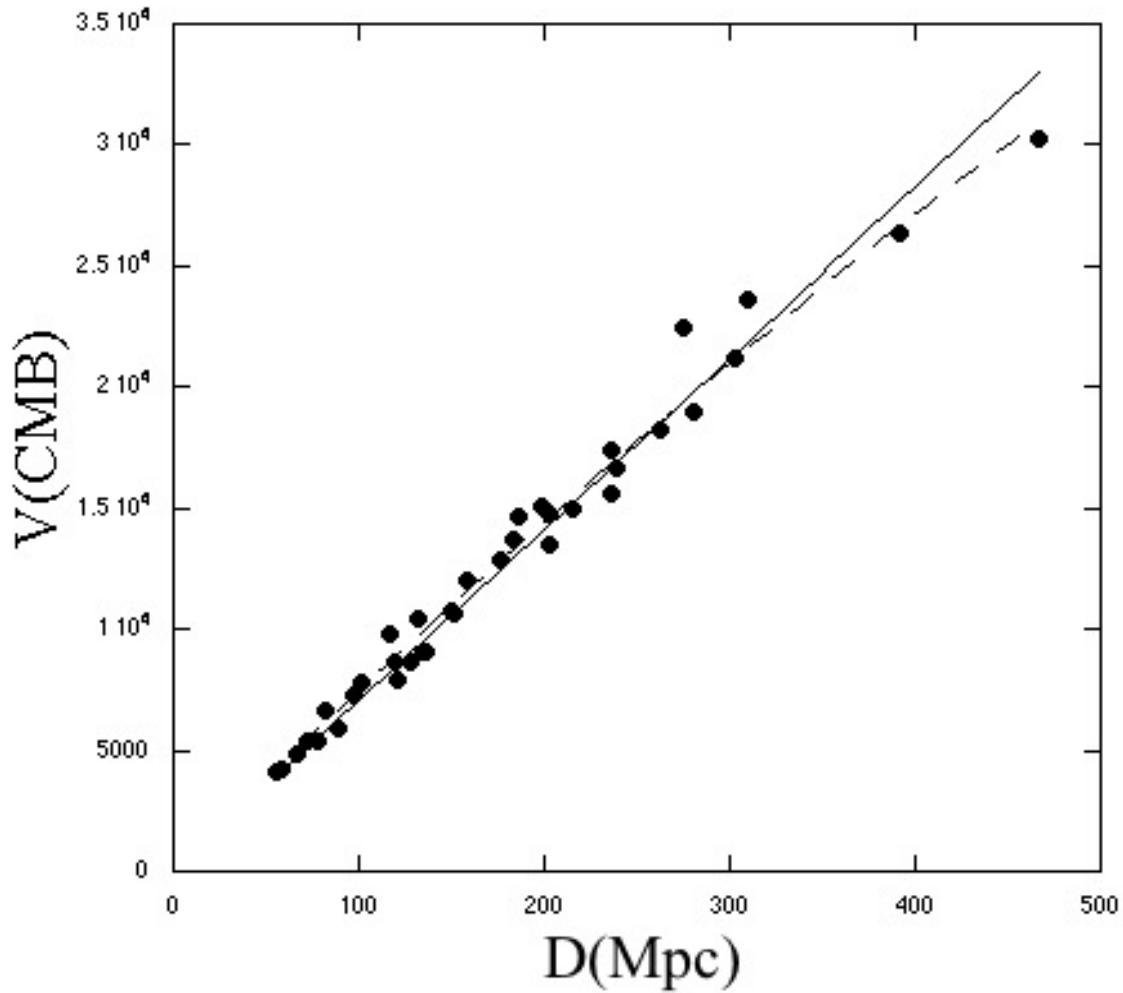

**Fig.1**. Velocity-distance plot. The data (full circles) are from the WMAP collaboration [2]. The full line is a fit according to Hubble's law [1] (**H=70.6 km s$^{-1}$Mpc$^{-1}$=2.310$^{-18}$s$^{-1}$**), while the dashed line takes into account second order corrections, resulting in a better reproduction of the data (**$H_0$=76.9 km s$^{-1}$Mpc$^{-1}$=2.410$^{-18}$s$^{-1}$ and $H_1$=0.022 km$^2$s$^{-1}$Mpc$^{-2}$ =2.310$^{-44}$m$^{-1}$s$^{-1}$**).

These quantities represent the limit of validity of our approach since $v(l_{MHL})=0$ from equation 6. Notice that the farthest distance measured and plotted in figure 1 is $D_{max}=1.4 \; 10^{25}$ m, well below the value in equation 7. Thus our considerations will be valid for distances of the order of $l_{MHL}$. Future data at larger distances might put more firm constraints. The apparently innocent modification of the HL has dramatic consequences, as we will demonstrate in a simple 'dust' model of the universe [3].

A spherical shell of mass m and radius r expands with velocity v. The total mass within the shell is given by M. Let us define the total energy per unit mass as: ε=E/m. It is given by [3]:

$$\varepsilon = \frac{1}{2}v^2 - \frac{GM}{r} = \frac{1}{2}(H_0 r - H_1 r^2)^2 - \frac{GM}{r} \quad (8)$$

Where we used the MHL, equation 6. We notice that the experimental result, equation 5, implies ε=0 and the universe is flat [2,3].

The appearance of higher order terms in equation 8, suggests the possibility that the universe might expand and finally reach an equilibrium position $r_{0s}$. This equilibrium position can be calculated from the first derivative of equation 8:

$$\varepsilon' = (H_0 r_{0s} - H_1 r_{0s}^2)(H_0 - 2H_1 r_{0s}) + \frac{GM}{r_{0s}^2} = 0 \quad (9)$$

Which gives:

$$\frac{GM}{r_{0s}^3} = -(H_0 - H_1 r_{0s})(H_0 - 2H_1 r_{0s}) > 0 \quad (10)$$

Since the mass must be positive, the range of equilibrium values, which satisfy equation [10], is rather restricted: $0.5 < r_{0s}/l_H < 1$. Assuming a flat universe, ε=0, gives the values:

$$\begin{cases} r_{0s} = \frac{3}{5} l_{MHL} = 6.25 \; 10^{25} m \\ t_{0s} = \frac{r_{0s}}{l_{MHL}} t_{MHL} = 2.52 \; 10^{17} s \\ v_{0s} = \frac{r_{0s}}{t_{0s}} = v_{MHL} = 2.5 \; 10^8 m/s \end{cases} \quad (11)$$

Which is the equilibrium value of the radius of the universe to be reached approximately at time $t_{0s}$. The first derivative of equation 9 calculated at $r_{0s}$ reveals that the equilibrium solution is a maximum thus unstable: after reaching $r_{0s}$, the fate of the universe will be determined by fluctuations: either collapse or expand forever!

From $r_{0s}$ we can estimate the total density of the universe (both dark and baryonic density) at time $t_{0s}$:

$$[\Omega_m]_{MHLS} = \frac{\rho_m}{\rho_c} = \frac{4}{25} \frac{H_0^2}{H^2} = 0.19 \quad (12)$$

Where we have used equations 2, 10 and 11 and we have taken into account the small difference between the Hubble's constants obtained using equations 1 and 6. The calculated value of the density is not too far away from the experimental value, equation 2, but it is definitely outside the quoted error. This suggests that the possibility of reaching an unstable equilibrium solution (even though at later times) might not be supported by the data. Other possibilities are that the universe has not expanded enough to reach $r_{0s}$ or that it has already expanded above $r_{0s}$ and it will never stop.

Before proceeding further, it is instructive to estimate the value of the DE using equation 8. Recall that in the ΛCDM approach, the energy per unit mass is given by:

$$\varepsilon_\Lambda = \frac{1}{2} v^2 - \frac{GM}{r} - \frac{1}{6} \Lambda c^2 r^2 = \frac{1}{2}(Hr)^2 - \frac{GM}{r} - \frac{1}{6} \Lambda c^2 r^2 \quad (13)$$

It is evident from equation 13 that the particular choice of the cosmological potential is to balance the kinetic part coming from the HL. Imposing $\varepsilon(r_{0s})=\varepsilon_\Lambda(r_{0s})$ gives:

$$[\Omega_\Lambda]_{MHLS} = \frac{\Lambda c^2}{3H^2} = 0.81 \quad (14)$$

The estimated value is again not too different from the experimental value given in equation 4 but definitely outside the experimental error as before. Notice that equation 5 is fulfilled in the model case. Before discussing under which conditions the data can be reproduced we must stress that in our model there is no need for *dark-energy*, thus Λ=0. This is an important consequence of the modification introduced in equation 6: *the extra energy needed to reproduce the experimental data is due to the kinetic component and not to the potential one as assumed in equations 3 and 13.*

6We can reverse the arguments above and, starting from the experimental value of the matter density, derive the value of the radius of the universe *at present time*. Assuming a stationary solution as before, we can derive the value of $r_{0s}$, which corresponds to the experimental density, equation 2. Using equations 10,12 and 2, it is a simple exercise to show that the experimental value of the matter density *does not support a stationary solution at present time even for* $\varepsilon \neq 0$. Alternatively, assuming a flat universe, $\varepsilon=0$, we can derive the matter density from equation 8 and put it equal to the experimental value. This gives solutions for the radius of the universe at *present time* (the age of the universe). Since equation 8 is quadratic in r we get two values:

$$r_0 = \begin{cases} 0.52 l_{MHL} \\ 1.48 l_{MHL} \end{cases} \qquad (15)$$

Notice that the smaller solution in equation 15 is not too much different from the stationary solution, equation 11. If this lower value is the actual radius of the universe, then the universe will expand and eventually reach the (unstable) equilibrium solution given in equation 11. The age of the universe corresponding to such $r_0$ is: $t_0 = \frac{r_0}{l_{MHL}} t_{MHL} = 1.31 \cdot 10^{17} s$, a little smaller than the equilibrium value $t_{0s}$.

The larger $r_0$ value in equation 15 is near the limits of validity of our approach given in 7. If a larger value than $r_{0s}$, would be supported from data, we could conclude that the universe has reached the unstable equilibrium position and, because of fluctuations, will continue to expand forever. The age of the universe corresponding to the larger $r_0$ value is $t_0=3.73 \cdot 10^{17}$ s, close to the quoted WMAP value $4.34 \cdot 10^{17}$ s [2,3], obtained from the inverse of the Hubble's constant H. Such an age value would support a universe expanding forever. Finally, using equation 15 to derive the DE density from equations 13 and 8, reproduces the experimental value given in equation 4 as expected.

It is instructive to recover the last result starting from the Friedmann equation from general relativity. Using the same formalism as in ref. [3] (equation (29.114)), the Friedmann equation, is written as:

$$\left[\left(\frac{1}{r}\frac{dr}{dt}\right)^2 - \frac{8}{3}\pi G(\rho_m + \rho_\Lambda)\right] r^2 = -kc^2 \qquad (16)$$

The latter equation is the equivalent of the Newtonian equation (8) where the role of the energy is taken by the space curvature *k*. The density due to relativistic particles has been neglected since its contribution is small. Since, $v(r) = \frac{dr}{dt}$, using the linear Hubble's law and the density values given in equations (2) and (4) gives k=0 i.e. a flat universe. If we now substitute the MHL, equation (6), into equation (16) and neglect the Dark Energy term, we obtain:

$$\left[H_0^2 \left(1 - \frac{r}{l_{MHL}}\right)^2 - \frac{8}{3}\pi G \rho_m\right] r^2 = -kc^2 \qquad (17)$$

Putting $r=r_0$, equation (15), and taking into account the small difference between H and $H_0$, gives $k=0$. Again the role of the DE has been taken by the correction in the kinetic part in agreement to the experimental data plotted in figure 1.

In conclusion, in this work we have shown that higher order corrections to Hubble's law have important consequences and in particular that there is no need for the dark energy, thus the cosmological constant $\Lambda=0$. The role of the dark energy is played by higher order corrections to the kinetic energy. We have estimated the value of the matter density (both baryonic and not) assuming the possibility that the universe will eventually reach an unstable equilibrium. The unstable equilibrium could be reached if the present value of the radius of the universe is about $5.4 \times 10^{25}$ m. On the other hand if the present value of the radius of the universe is larger than $6.25 \times 10^{25}$ m, i.e. above the equilibrium point, then the universe will expand forever. Because of the sensitivity of our results to the data we do not exclude the possibility that further measurements of the velocity of recessions of galaxies as function of the distance will give more constraints on the polynomial expansion proposed in this work. In particular we need measurements at smaller as well as larger distances than reported in figure 1.

In loving memory of my son Luca (1992-2012).